\documentclass[preprint,onecolumn,nofootinbib]{revtex4}
\pdfoutput=1
\usepackage[colorlinks=true,linkcolor=blue,urlcolor=blue,filecolor=black,citecolor=red,pdfstartview=FitV,pdftitle={},pdfsubject={},pdfkeywords={},pdfpagemode=None,bookmarksopen=true]{hyperref}
\usepackage{graphicx}
\usepackage{amsmath}
\usepackage{amsfonts}
\usepackage{amssymb,ulem}
\usepackage{color}%
\usepackage{dcolumn}
\usepackage{MnSymbol,wasysym}
\usepackage{braket}

\usepackage{eurosym}
\usepackage{calrsfs}
\usepackage[usenames,dvipsnames,svgnames]{xcolor}
\usepackage{color}%

\begin{document}
\title{Joule-Thomson expansion in AdS black holes with momentum relaxation}
\author{Adolfo Cisterna}
\email{adolfo.cisterna@ucentral.cl}
\affiliation{Vicerrector\'ia Acad\'emica, Toesca 1783, Universidad Central de Chile, Santiago, Chile}
\author{Shi-Qian Hu}
\email{mx120170256@yzu.edu.cn}
\author{Xiao-Mei Kuang}
\email{xmeikuang@yzu.edu.cn}
\affiliation{Center for Gravitation and Cosmology, College of Physical Science and Technology, Yangzhou University, Yangzhou, 225009, China}

\date{\today }
\begin{abstract}

The inner structure of realistic materials make them exhibit momentum relaxation. In this paper we study the Joule-Thomson effect on AdS black holes in which translational invariance is broken by two methods: First by considering planar black holes in general relativity supported by scalar fields with a linear dependence on the horizon coordinates and secondly by considering black holes in massive gravity models in which momentum relaxation is obtained by breaking the bulk diffeomorphism invariance of the theory. In contrast with black holes studied so far, for both theories it is possible to obtain inversion curves with two branches reproducing the behavior of Van der Waals fluids. Moreover in the specific case of the massive gravity model we show that black holes can heat up when crossing the inversion curve.
\end{abstract}

\maketitle

\tableofcontents

\section{Introduction}

From the very beginning black holes were described as extreme classical objects that absorb every kind of matter and energy without leaving anything out. Basically regarded as bald objects \cite{Ruffini:1971bza,Bekenstein:1998aw} they were supposed to be described just by few parameters; mass, angular momentum and electromagnetic charges. With the advent of quantum field theory, particularly in the context of curved spacetimes, it was demonstrated the fundamental relationship existing between the area of black holes and their entropy \cite{entropy1} and that they possess a temperature related with its surface gravity \cite{tem1}. Moreover it was shown that black holes emit radiation resembling the spectrum of black bodies. All these considerations led to the development of black hole thermodynamics \cite{termo1}, constituting the first successful semiclassical description of gravitational phenomena, a deep insight into the understanding of a possible quantum description of the gravitational interaction.

When considering black holes in Anti-de Sitter spacetimes black hole thermodynamics becomes particularly interesting. The AdS/CFT correspondence defines a duality between gravitational theories on anti-de Sitter spacetimes in $(D+1)-$dimensions and conformal field theories in $D-$dimensions \cite{Maldacena}. In this context black holes in the presence of a negative cosmological constant admits a dual description given by thermal states in a conformal field theory. Hawking and Page \cite{HPT} demonstrated that AdS spacetimes suffer a phase transition from the AdS background state to large Schwarzschild AdS black holes for a critical temperature. Phase transition that was demonstrated to be dual to a confinement/deconfinement phase transition in the free energy of the dual field theory quark/gluon plasma \cite{Witten:1998qj,witten2}. Many applications of these ideas that combine black hole thermodynamics and AdS/CFT duality have been developed during the last decade providing a deeper understanding of the interplay between gravity and quantum physics in the context of condense matter physics, \cite{hart1,hart2}, the loss information paradox \cite{info1}, quantum chromodynamics \cite{cromo1}, to mention few examples.

When considering black hole thermodynamic for AdS black holes the cosmological constant parameter, $\Lambda$, is considered as a fixed parameter introduced in the action and does not appears in the first law of black hole thermodynamics. This ensures that we are comparing thermodynamical ensembles for solutions exhibiting the same asymptotic behavior, by fixing the AdS background.

Despite its original constant nature, it is interesting to explore how black hole thermodynamic is modified when considering the cosmological constant, $\Lambda$, as a thermodynamical variable \cite{Kastor:2009wy,lam2}. In fact, the cosmological constant, from a  perfect fluid point of view incorpores the notion of Pressure, through the relation
\begin{equation}
 P=-\frac{\Lambda}{8\pi}=\frac{(d-1)(d-2)}{16\pi l^2}
\end{equation}
where $d$ represents the dimension of the spacetime and $l$ the AdS radius. This allows to obtain a more physical interpretation of what the volume of a black hole should be \cite{lam3,lam4}, thermodynamical volume defined by \footnote{It has been stressed \cite{lam3} that the thermodynamical volume seems to be equal or more than the corresponding Euclidean volume associated with the area/entropy. This implies that black holes are more efficient when storing information. This result is known as the reverse isoperimetric inequality (RII) and has been analyzed for several black hole solutions \cite{exam1,exam2,exam3,exam4,Feng:2017jub}.}
\begin{equation}
V=\left(\frac{\partial{M}}{\partial{P}}\right)_{S,J,Q}.
\end{equation}
This permits the inclusion of the pressure-volume term of everyday thermodynamic $PV$ into the first law of black hole thermodynamic \cite{Dolan:2012jh}, term that does not appears otherwise. Moreover provides the correct Smarr relation for AdS black holes delivering a concrete expression relating extensive and intensive thermodynamic variables. In this case the mass $M$ of the spacetime must be interpreted as the enthalpy of the thermodynamical system \cite{Kastor:2009wy}. Novel new phenomenology is obtained, new phase transitions like Van der Waals liquid-gas phase transitions \cite{mann1,mann2}, existence of triple points like the one encountered in the phase diagram of water \cite{Kubiznak:2016qmn}, heat engines black hole analogous, just to mention few applications \cite{Johnson:2014yja,Hennigar:2017apu,Fang:2017nse,Hu:2019wwn}. The subject has been dubbed generalized black hole thermodynamic or black hole chemistry \cite{Kubiznak:2016qmn}.

An interesting classical thermodynamical effect is the so-called Joule-Thomson effect, also known as Joule-Thomson expansion \cite{Ref-JT}. This effect deals with the change of temperature of a gas or fluid when it is expanded adiabatically by using a valve. In fact this adiabatic expansion can be performed in several ways. The Joule-Thomson effect takes place when the thermodynamical process occurring during the expansion is irreversible and enthalpy remains constant. The change of temperature is measured by the Joule-Thomson coefficient $\mu_{JT}$ which can be either positive or negative depending if the fluid is cooling or heating, respectively.

By working on the context of generalized black hole thermodynamic recently the Joule-Thomson effect have been studied for first time by \"{O}kc\"{u} and Ayd{\i}ner \cite{Okcu:2016tgt}, in particular for the case of Reissner-Nordstrom black holes in anti-de Sitter spacetimes.

As we have stated previously, the Joule-Thomson expansion deals with the change of temperature of the fluid under expansion in an isenthalpic process. This change is quantitatively expressed by the sign of the Joule-Thomson coefficient defined by \cite{Ref-JT}
\begin{equation}
\mu_{JT}=\left(\frac{\partial T}{\partial P}\right)_{H}.
\end{equation}
We observe that by computing this coefficient it is possible to determinate when heating or cooling is taking place. Even if pressure is always decreasing the change of temperature can be either positive or negative. When $\mu_{JT}$ goes to zero it is possible to defined the inversion temperature $T_{i}$, the particular point in the gradient of temperature of the black hole for which the system change from cooling to heating or vice versa. In the same manner it is defined the inversion pressure $P_{i}$. Then $(P_i,T_i)$ gives inversion transition point. By making use of the generalized first law of black hole thermodynamic and taking into account the isenthalpic nature of the process, it is possible to define the Joule-Thomson coefficient in term of the volume and heat capacity at constant pressure \cite{Okcu:2016tgt},
\begin{equation}
\mu_{JT}  =\left( \frac{\partial T}{\partial P}\right) _H=\frac{1}{C_P}\left[ T\left( \frac{\partial V}{\partial T}\right) _P-V\right], \label{eq-mu}
\end{equation}
where $C_P=T\left(\frac{\partial S}{\partial T}\right)_P$ is the heat capacities at
constant pressure.
This definition has the advantage that allows to define easily the inversion temperature
\begin{equation}
T_{i}=V\left( \frac{\partial T}{\partial V}\right)_P
\end{equation}
which will provide heating and cooling regions in the $T-P$ plane.

The study of Joule-Thomson expansion has been generalized  for several black hole solutions including Quintessence AdS black hole \cite{Ghaffarnejad:2018exz}, arbitrary dimensional charge AdS black hole \cite{Mo:2018rgq}, Kerr-AdS black holes \cite{Okcu:2017qgo}, Kerr-Newman-AdS black holes\cite{Zhao:2018kpz},Gauss-Bonnet AdS black holes \cite{Lan:2018nnp}, Lovelock gravity \cite{Mo:2018qkt} and nonlinear electrodynamic gravity \cite{Kuang:2018goo} to mention few examples.

It is known that real materials exhibit momentum dissipation, namely, that momentum is not continuously conserved. This implies that resistivity of materials has a non-vanishing value providing for finite electrical conductivities. When making use of the tools of the gauge/gravity duality to study condense matter systems in term of their gravitational dual it is not straightforward to include momentum dissipation. At this respect to well-known strategies to produce momentum dissipation are the inclusion of matter fields that breaks translation invariance in the dual field theory, as it is for example the case of the scalar fields that depend linearly on the horizon coordinates \cite{Andrade:2013gsa}, and the case of massive gravity theories which present a broken diffeomorphism invariance in the bulk \cite{Vegh:2013sk}.

In this paper we present the study of the Joule-Thomson effect for two models presenting momentum dissipation: $(i)$ the Einstein-Maxwell-scalar theory where the scalar fields act as spatial-dependent sources breaking the Ward identity so that the momentum is not conserved in the dual theory \cite{Andrade:2013gsa}; $(ii)$ massive gravity theory where the momentum dissipation in the dual theory is implemented by breaking the diffeomorphism invariance in the bulk \cite{Vegh:2013sk}.  We will work in the units with $G=c=\hbar=k_B=1$.

 The paper is organized as follows. Sec. II is designated to analyze the Joule-Thomson expansion in the context of Einstein-Maxwell-scalar theory. This is done for scalar fields presenting a standard kinetic term but also for the case in which the kinetic term is modified by the so-called k-essence term \cite{Baggioli:2014roa,Cisterna:2017jmv}. Sec. III is devoted to the analysis of the Joule-Thomson expansion in the context of massive gravity theory. Finally we conclude in Sec. IV.
\section{Joule-Thomson expansion in Einstein-Maxwell-scalars theory }
\subsection{Black holes in Einstein-Maxwell-scalars theory}
The Einstein-Maxwell-scalars gravity theory with scalar fields was first proposed in \cite{Andrade:2013gsa} by homogeneously distributing $2$ massless scalar fields along the horizon coordinates. The action principle in four dimensions reads
\begin{equation}
S=\frac{1}{16\pi G}\int \! d^4x \sqrt{-g} \left(R-2\Lambda -\frac{1}{4}F_{\mu\nu}F^{\mu\nu}-\frac{1}{2}\sum_{i=1}^{2}(\partial\psi_i)^2\right)\ ,
\label{eq:action}
\end{equation}
where the cosmological and the AdS radius are related by $l^2=-\frac{3}{\Lambda}$.

By setting the scalar fields to depend on the $2$ dimensional spatial coordinates and by considering a spacetime with a planar base manifold the Klein-Gordon equation for the scalars is easily integrated, yielding
\begin{equation}
\psi_I=\beta_{Ia}x^a.
\end{equation}
Subsequently one finds that the action admits the following charged black hole solution
 \begin{eqnarray}\label{eq-metric}
&&ds^2=-f(r)dt^2+\frac{1}{f(r)}dr^2+r^2dx^a dx^a,~~~~~A=A_t(r) dt,~~\mathrm{with}\nonumber\\
&&f(r)=\frac{r^2}{l^2}-\frac{\beta^2}{2}-\frac{m}{r}+\frac{q^2}{r^{2}}, ~~~A_t=\left(1-\frac{r_h}{r}\right)\frac{2q}{r_h}
 \end{eqnarray}
where the index $a$ goes $a=1,2$, and the horizon $r_h$ satisfies $f(r_h)=0$.
It is worthwhile to point out that the scalar fields in the bulk source a spatially dependent field theory with momentum relaxation, which is dual to the homogeneous and isotropic black hole \eqref{eq-metric} \footnote{From a geometric point of view the parameter $\beta$ induces and effective negative curvature scale on the horizon, resembling the causal structure of hyperbolic black holes. This was first observed in \cite{Bardoux:2012aw}.}. The linear coefficient $\beta$ of the scalar fields somehow can be considered to describe the strength of the momentum relaxation in the boundary theory \cite{Andrade:2013gsa} \footnote{Recently scalars fields of this type have been used to construct exact anti-de Sitter homogeneous black branes \cite{Kuang:2017cgt,Cisterna:2018hzf} and black strings \cite{Cisterna:2017qrb}.}.

The mass and the charge of this black hole can be found by means of the Hamiltonian analysis or by using the holographic renormalization approach developed in \cite{Bardoux:2012aw, Andrade:2013gsa} \footnote{It is important to note that under this approach the axion fields appears explicitly in the first law and that their conjugate variables are interpreted as magnetic susceptibilities. It is fair to say that this interpretation is still a subject of debate by following the lines of the proposal \cite{Gibbons:1996af} and its generalization \cite{Astefanesei:2018vga} in which the first law does not contains extra term coming from possible hairs parameters, in such a case, scalar fields that contribute non-trivially to the mass of the solution.}.
In there has been shown that
the mass and charge of the black hole are connected with the parameters $m$ and $q$ as
\begin{eqnarray}\label{eq-MQ}
M=\frac{\mathcal{V}_2}{8\pi}m,~~~\mathrm{and}~~~Q=\frac{\mathcal{V}_2}{8\pi}q
\end{eqnarray}
where $\mathcal{V}_2$ is the volume of the $2$ dimensional flat space and we will set it to be $1$. On the other hand, by identifying the period of the Euclidean time in order to avoid conical singularities, the temperature of the black hole is given by
\begin{equation}\label{eq-T}
T=\frac{f'(r_h)}{4\pi}=\frac{1}{4\pi}\left(\frac{3 r_h}{l^2}-\frac{\beta ^2}{2r_h}-\frac{q^2}{r_h^{3}}\right),
\end{equation}
and the entropy is obtained by the area law as
\begin{equation}\label{eq-S}
S=\frac{r_h^{2}}{4}.
\end{equation}

\subsection{Joule-Thomson expansion}

 We shall apply the previous solution to the study of the Joule-Thomson expansion. To do so we consider the thermodynamical analysis of the Einstein-Maxwell-scalar theory provided in \cite{Fang:2017nse} in which the cosmological constant is consider as a thermodynamical parameter susceptible to variations. The first law of black hole mechanics/thermodynamic under this approach has been properly identified by means of the hamiltonian formalism in \cite{Kastor:2009wy}. It is important to stress that even if this upgrade of the original parameter nature of the cosmological constant is done straightforwardly, this can be done following the approach developed in the seminal paper \cite{Henneaux:1984ji} where the cosmological constant is given by the inclusion of a three-index antisymmetric Abelian gauge field, which once integrated promote the cosmological constant to a canonical variable. Nevertheless it is fair to establishes that this is a subject of debate and that so far this treatment can be considered as a proposal more than a fundamental conjecture, indeed it offers many applications in black hole thermodynamics in close analogy with common thermodynamics but it has some subtleties when applying the full AdS/CFT dictionary.

\begin{figure}
\center{
\includegraphics[scale=0.7]{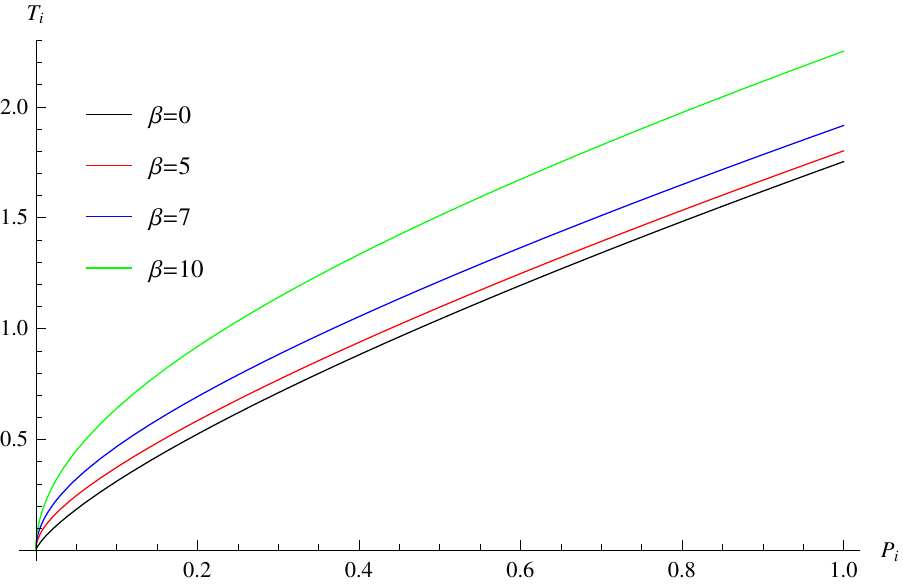}
\caption{\label{fig-Pi-Ti-axion}Inversion curves $T_i-P_i$ for heating and cooling processes for different values of $\beta$. We set $q=5$.}}
\end{figure}

 As it is known the pressure is given by
\begin{equation}\label{eq-Pl}
P=-\frac{\Lambda}{8\pi}=\frac{3}{8\pi l^2}.
\end{equation}
Making use of this result into the definition of mass,  the mass of the black hole can be rewritten as
\begin{equation}
M=\frac{P r_h^3}{3}+\frac{q^2}{8\pi r_h}-\frac{\beta^2 r_h}{16\pi}
\end{equation}
which is taken as the enthalpy $H$ of the system.
On the other hand, let us use pressure \eqref{eq-Pl} into the expression for the temperature \eqref{eq-T}, then
\begin{equation}\label{eq-P}
P=\frac{ T}{2r_h} +\frac{\beta^2}{16\pi r_h^2}+\frac{q^2}{8\pi r_h^{4}}\ .
\end{equation}
In this manner we obtain our black hole equation of state. The thermodynamical volume is the conjugate variable of the pressure, then
\begin{equation}\label{eq-V}
V=\left(\frac{\partial M}{\partial P}\right)_{Q,S}=\frac{r_h^3}{3}.
\end{equation}
With these ingredients at hand we use the definition of the Joule-Thomson coefficient   (\ref{eq-mu}), obtaining
\begin{equation}
\mu_{JT}=\frac{2 r_h \left(r_h^2 \left(\beta ^2+16 \pi  P r_h^2-24 \pi  T  r_h\right)+6 q^2\right)}{-48 \pi  P r_h^4+3 \beta ^2 r_h^2+6 q^2}.
\end{equation}
As was stated before, the point $\mu_{JT}$=0 defines the inversion temperature $T_{i}$, which in our case reads
\begin{equation}\label{eq-Ti}
T_i=\frac{16 \pi  P_i r_h^4+\beta ^2 r_h^2+6 q^2}{24 \pi  r_h^3}
\end{equation}
with the corresponding pressure $P_i$. As we see this expression depends apart from the inversion pressure on the horizon of the black hole. Nevertheless $r_h$ can be obtained from the temperature relation, relation that $T_i$ and $P_i$ must also satisfy, then the only positive root is
\begin{equation}\label{eq-rh}
r_h=\frac{1}{4} \sqrt{\frac{\sqrt{\beta ^4+96\pi  P_i q^2}}{\pi  P_i}+\frac{\beta ^2}{\pi  P_i}}.
\end{equation}
Substituting \eqref{eq-rh} into \eqref{eq-Ti}, we obtain the analytical relation between the inversion temperature and pressure
\begin{equation}\label{eq-PiTi}
T_i=\frac{\beta ^4+\beta ^2 \sqrt{\beta ^4+96 \pi  P_i q^2}+64 \pi  P_i q^2}{2 \sqrt{\pi } P_i \left(\frac{\beta ^2+\sqrt{\beta ^4+96 \pi  P_i q^2}}{P_i}\right)^{3/2}}.
\end{equation}
From this last equation we observe some analytical properties of the inversion temperature. For the case without momentum relaxation, i.e, $\beta=0$, we have $T_i=\frac{2 \sqrt[4]{\frac{2}{\pi }} P_i^{3/4} \sqrt{q}}{3^{3/4}}$ which is proportional to $P_i^{3/4}$ with fixed $q\neq 0$. This shows that for planar charged AdS black holes
the minimum temperature goes to zero when the inversion pressure tends to zero, contrary to the case in which the horizon is spherical. Moreover there is no inversion temperature for uncharged solutions. On the other hand when $q=0$, but we have momentum relaxation we obtain that $T_i=\frac{\mid\beta\mid\sqrt{P_i}}{2 \sqrt{2 \pi }}$ which is shifted from zero in contrast with the standard uncharged
AdS black hole \cite{Okcu:2016tgt}.
We show the explicit relation of inversion curve for different $\beta$ in figure \ref{fig-Pi-Ti-axion}.

There is only one branch of inversion curves as that studied in \cite{Okcu:2016tgt,Mo:2018qkt,Lan:2018nnp,Okcu:2017qgo,Mo:2018rgq,Kuang:2018goo}, which differ from the one obtained for Van der Waals fluids.
As momentum increases, the curve is higher. This effect is similar to the one produced by the electric charge. So the momentum relaxation enhances the inverse curve.

\begin{figure}
\center{
\includegraphics[scale=0.7]{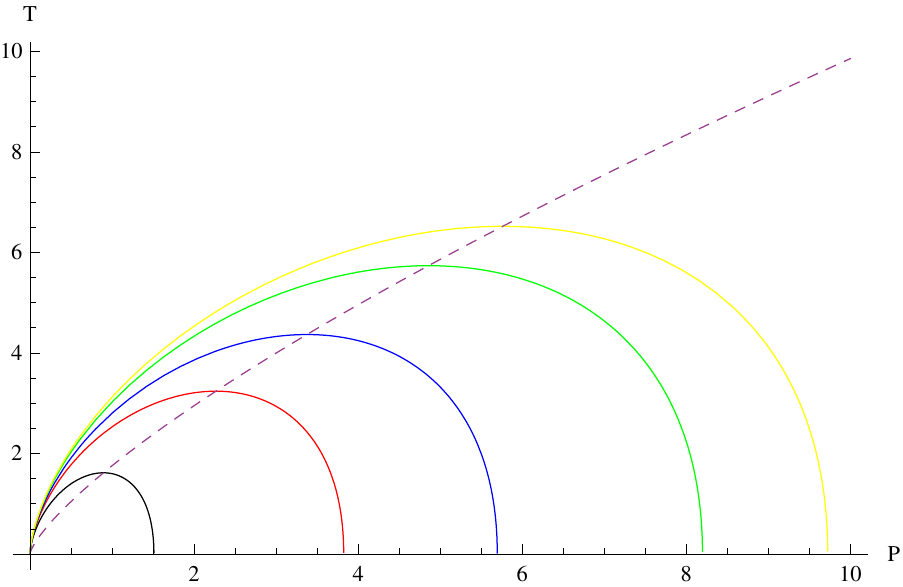}
\caption{\label{fig-P-T-axion}The isenthalpic curves. The mass are $1.4,1.8,2,2.2,2.3$ from inner to outer plots. We set $q=5$ and $\beta=2$.}}
\end{figure}

\begin{figure}
\center{
\includegraphics[scale=1]{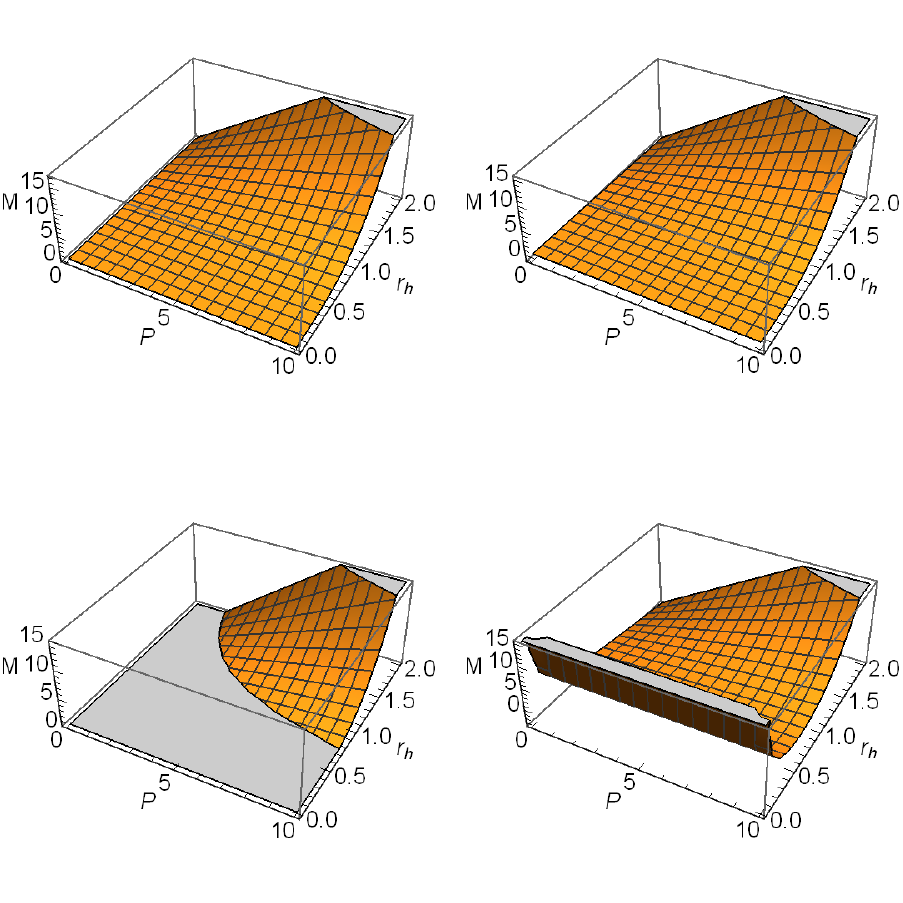}
\caption{\label{fig-3D} Relations of mass, event horizon and pressure. The parameters are $\beta=q=0; \beta=5, q=0 ;\beta=10, q=0$ and $\beta=10, q=5$, respectively.}}
\end{figure}

We turn to study the isenthalpic curves with constant mass/enthalpy in the $T-P$ plane. The results for $q=5$ and $\beta=2$ is displayed in figure \ref{fig-P-T-axion} where the solid lines are isenthalpic curves while the dashed line corresponds to the inverse curve. For the isenthalpic process in the left side of inversion curve, the temperature increase as the pressure, so $\mu_{JT}>0$ denotes a warming process while in the right side, $\mu_{JT}<0$ denotes a cooling process.
Furthermore, in figure \ref{fig-3D} we plot the mass in terms of the horizon and pressure by fixing $\beta$ and $q$.  We can only study the Joule-Thomson expansion for a positive event horizon, so that isenthalpic curves are real.


The previous studies can be generalized to the case in which the kinetic term for the scalar fields can have non-linear contributions. Such a case can be implemented by the so-called k-essence models \cite{ArmendarizPicon:1999rj}, in which the mentioned kinetic term are generalized to be a function $P(\psi, (\partial\psi)^2)$. A simple case contained on this setup is the one in which the scalar fields apart from the standard kinetic term possess a kinetic nonlinear contribution given by higher powers of the kinetic term. The action was studied in \cite{Baggioli:2014roa,Cisterna:2017jmv} and is given by
\begin{equation}
S=\frac{1}{16\pi}
\int\sqrt{-g}\left(R-2\Lambda-\frac{1}{4}F_{\mu\nu}F^{\mu\nu}-\sum_{i=1}^{2}(X_i+\gamma X_i^k)\right)d^4x\,,
\label{action}
\end{equation}
where  $X_{i}=\frac12\nabla^{\mu}\psi_{i}\nabla_{\mu}\psi_{i}$ with $i=1,2$. $\psi_i$ are massless scalar field. The above action
goes back to that for the minimally coupled Einstein-Maxwell-scalars gravity studied in \cite{Andrade:2013gsa} just by setting $\gamma=0$. The exact black hole solution for this theory was found in \cite{Cisterna:2017jmv}
\begin{eqnarray}
ds^2&=&-f(r)dt^2+\frac{dr^2}{f(r)}+r^2(dx_1^2+dx_2^2),\,\label{metriccharged}\\
f(r)&=&\frac{r^2}{l^2}-\frac{2m}{r}-\frac{\lambda^2}{2}+\gamma\frac{\lambda^{2k}}{2^k(2k-3)}r^{2(1-k)}+\frac{q^2}{
r^2}, \label{metricf}
\end{eqnarray}
with unchanged matter fields respect to the previous analyzed solution
\begin{eqnarray}
\psi_1&=&\beta x_1,\ \psi_2=\beta x_2\ , \label{axions}\\
A&=&\left(\rho_0-\frac{2q}{r}\right)dt\,.\label{At}
\end{eqnarray}
The extended thermodynamics of the above solution was studied by us in \cite{Hu:2019wwn}.
The Hawking temperature and the entropy of this black hole are given by \begin{equation}\label{Temp}
T =\frac{f'(r_h)}{4\pi}=2 P r_h-\frac{q^2}{4 \pi  r_h^3}-\frac{\beta^2}{8 \pi  r_h}-\frac{\gamma  2^{-k-2} \beta ^{2 k} r_h^{1-2 k}}{\pi },
\end{equation}
\begin{equation}\label{eq-S2}
S =\frac{ r_h^2}{4}.
\end{equation}
We see that the temperature is modified by $\gamma$ while the entropy is the same as \eqref{eq-S}.
The mass of the black hole and the charge are connected with $m$ and $q$ by means of
\begin{equation}\label{MQ}
M=\frac{1}{16\pi}\left(\frac{\gamma  2^{1-k} \beta ^{2 k} r_h^{2 (1-k)+1}}{2 k-3}+\frac{16}{3} \pi  P r_h^3+\frac{2q^2}{ r_h}-\beta ^2 r_h\right),
~~Q=\frac{q}{8\pi}
\end{equation}
where we have used the definition of pressure. The mass is computed by means of the Euclidean approach, namely, the partition function for a thermodynamical ensemble is identified with the Euclidean path integral in the saddle point approximation around the classical Euclidean solution. The detailed computations are shown in detail in \cite{Cisterna:2017jmv} where the explicit realization of the first law of black hole thermodynamic for all values of $k$ is shown \footnote{This computation is based on the hamiltonian analysis performed in \cite{Bardoux:2012aw}.}.

The charge and  the thermodynamical volume as the conjugation of the pressure are the same as \eqref{eq-MQ} and \eqref{eq-V}, respectively.

\begin{figure}
\center{
\includegraphics[scale=0.7]{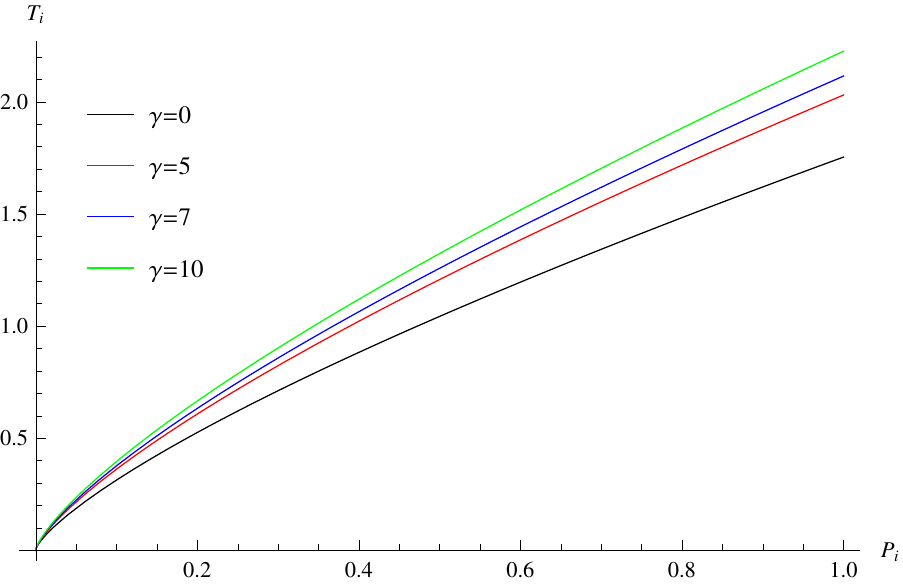}
\caption{\label{fig-Pi-Ti-gamma}Inversion curves $T_i-P_i$ for heating and cooling processes for different values of $\gamma$. We set $q=5$ and $\beta=2$.}}
\end{figure}

In order to analyze quantitatively the effect of the k-essence contribution on the Joule-Thomson expansion we will consider the $k=2$ case.
By following the same strategy followed previously we obtain that for the inverse curve the horizon should satisfy the equation
\begin{equation}\label{eqrh}
3 \gamma  \beta ^4-32 \pi  r_h^4 P_i+4 \beta ^2 r_h^2+12 q^2=0.
\end{equation}
By allowing only positive values for $\gamma$, no phantom contributions, there is only one positive root for $r_h$
\begin{equation}\label{eq-rh2}
r_h=\frac{1}{4} \sqrt{\frac{\beta ^2}{\pi  P_i}+\frac{\sqrt{\beta ^4+24 \pi  \beta ^4 \gamma  P_i+96 \pi  q^2 P_i}}{\pi  P_i}}.
\end{equation}
Then, our inversion temperature is related with the inverse pressure by
\begin{equation}\label{eq-PiTi2}
T_i=\frac{ (\sqrt{\sqrt{24 \pi  P_i \left(\gamma  \beta ^4+4 q^2\right)+\beta ^4}+\beta ^2)P_i} \left(\beta ^4+16 \pi  P_i \left(\beta ^4 \gamma +4 q^2\right)+\beta ^2 \sqrt{\beta ^4+24 \pi
   P_i \left(\beta ^4 \gamma +4 q^2\right)}\right)}{2 \sqrt{\pi } \left(\beta ^2+\sqrt{\beta ^4+24 \pi  P_i \left(\beta ^4 \gamma +4 q^2\right)}\right){}^2}.
\end{equation}
The inverse curve for different values of the $\gamma$ coupling are shown in figure \ref{fig-Pi-Ti-gamma}.
Similarly, the higher coupling also enhances the inverse curve.

\begin{figure}
\center{
\includegraphics[scale=1]{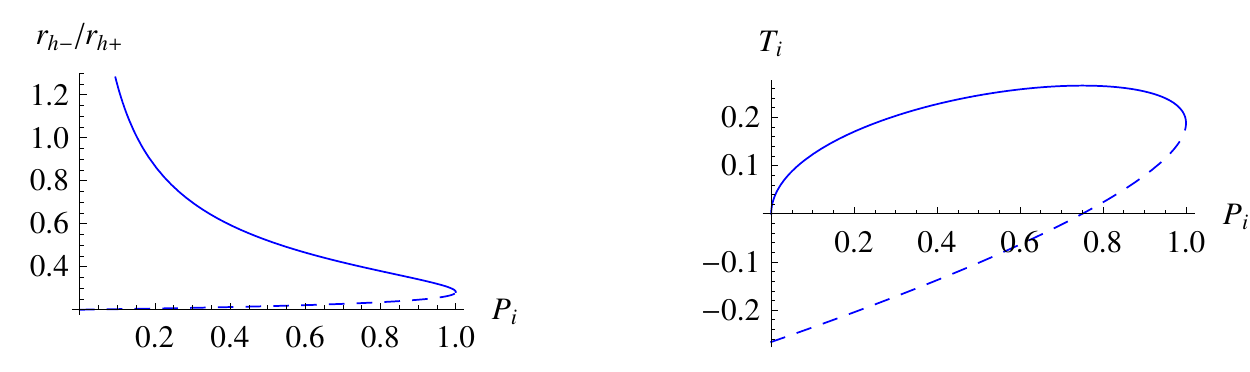}
\caption{\label{fig-Pi-Ti-gamma2}Left: The possible positive  horizon VS the inversion pressure. Right: Inversion curves $T_i-P_i$ for heating and cooling processes for $\gamma=-\frac{25}{4}-\frac{1}{24\pi}$. We set $q=5$ and $\beta=2$.}}
\end{figure}

Now we analyze the case in which $\gamma$ can take negative values. When considering this phantom contribution equation \eqref{eqrh} possesses two positive roots defined by
\begin{equation}\label{eq-rh2}
r_{h_-}=\frac{1}{4} \sqrt{\frac{\beta ^2}{\pi  P_i}-\frac{\sqrt{\beta ^4+24 \pi  \beta ^4 \gamma  P_i+96 \pi  q^2 P_i}}{\pi  P_i}},~r_{h_+}=\frac{1}{4} \sqrt{\frac{\beta ^2}{\pi  P_i}+\frac{\sqrt{\beta ^4+24 \pi  \beta ^4 \gamma  P_i+96 \pi  q^2 P_i}}{\pi  P_i}}.
\end{equation}
It is straightforward to compute when $-\frac{4q^2}{\beta^4}>\gamma>-\frac{1}{24\pi P_i}-\frac{4q^2}{\beta^4}$, both $r_{h_-}$ and $r_{h_+}$ are real positive roots, otherwise, only $r_{h_+}$ is positive. Substituting the two positive solution into the equation for $\mu_{JT}=0$, we can obtain two branches of the inversion curve. We show the possible positive root  and the inversion curve with the related  two branches  in figure \ref{fig-Pi-Ti-gamma2}.  However, with the same parameters, we found that the mass of black hole is negative for both  $r_{h_-}$ and $r_{h_+}$, which implies that in this case, the Joule-Thomson expansion breaks down for negative $\gamma$.  This is reasonable because negative $\gamma$ involves instability.

\section{Joule-Thomson expansion in massive gravity }
\subsection{Planar black holes in massive gravity}
Now we turn to the study of the Joule-Thomson expansion in the context of a massive gravity theory. We shall focus on four dimensional black holes with planar horizon. The action of the four-dimensional massive gravity we are considering is given by \cite{Vegh:2013sk}
\begin{equation}
S=\frac{1}{16\pi}\int
d^{4}x\sqrt{-g}\Big[R+\frac{6}{l^2}-\frac{1}{4}F^2+m_g^2\sum_{i=1}^4c_i\mathcal{U}_i(g,f)\Big], \label{1}
\end{equation}
where $m_g$ is the parameter controlling the massive term. We know that the presence of impurities in realistic materials involves that
momentum is not conserved which brings in finite DC conductivity. Modeling
systems with the use of  translationally invariant quantum field theories always have problems
unless the effects of momentum dissipation are incorporated. Fortunately, in the framework of holography, several methods employed to
include momentum dissipation and extract the DC conductivity have been proposed. Massive gravity with the action \eqref{1} proposed by Vegh \cite{Vegh:2013sk} is an alternative approach considering momentum dissipation holographically which provides an effective bulk description of a theory that does not conserve
momentum without using any specific mechanism.
In the action, in contrast with Einstein gravity, the last terms represent massive potentials associated with the graviton mass which breaks the diffeomorphism invariance in the bulk producing momentum relaxation in the dual boundary theory. The study of ghost-free massive gravity can be seen in \cite{Hassan:2011hr,Hassan:2011tf}.

In the action \eqref{1}, The couplings $c_i$ are series of constants with dimension while $f$ and $\mathcal{U}_i$ denote the reference metric and symmetric polynomials of the eigenvalue of the ($4$)$\times$($4$) matrix $\mathcal{K}^{\mu}_{~\nu}\equiv\sqrt{g^{\mu\alpha}f_{\alpha\nu}}$, respectively.  $\mathcal{U}_i$ have the forms
\begin{eqnarray}
\mathcal{U}_1&=&[\mathcal{K}],~~~\mathcal{U}_2=[\mathcal{K}]^2-[\mathcal{K}^2],\nonumber\\
\mathcal{U}_3&=&[\mathcal{K}]^3-3[\mathcal{K}][\mathcal{K}^2]+2[\mathcal{K}^3],\nonumber\\
\mathcal{U}_4&=&[\mathcal{K}]^4-6[\mathcal{K}^2][\mathcal{K}]^2+8[\mathcal{K}^3][\mathcal{K}]+3[\mathcal{K}^2]^2-6[\mathcal{K}^4] \label{2}
\end{eqnarray}
where $[\mathcal{K}]=\mathcal{K}^{\mu}_{~\mu}$ and the square root in $\mathcal{K}$ can be interpreted as $(\sqrt{\mathcal{K}})^{\mu}_{~\nu}(\sqrt{\mathcal{K}})^{\nu}_{~\lambda}=\mathcal{K}^{\mu}_{~\nu}$.  It is noted that even though self-consistent massive gravity theory \emph{may} require $c_i$ to be negative if $m_g^2>0$, the stability of fluctuations of fields deserves further analysis because the fluctuations of some fields with negative mass square could still be stable if the mass square obeys their related Breitenlohner-Freedman bounds\cite{Cai:2014znn}. Thus, in order to see more possible effect of massive terms on the JT expansion, here we do not take care of the sign of $c_i$ and  will consider both positive and negative values.

The static planar black hole solution of the above action yields~\cite{Vegh:2013sk,Cai:2014znn}
\footnote{We note that with the choice of degenerate background metric $f_{xx} = f_{yy} = c_0$ and
all other components vanishing, the action \eqref{1} preserves diffeomorphism invariance in the $(r; t)$ directions, but not $(x; y)$ directions. Therefore, the boundary dual theory does not have conserved momentum currents even though energy currents is conserved. Moreover,
the solution \eqref{6} is not $AdS$ due to the non-vanishing graviton mass and the degenerate choice of
the reference metric $f_{ij}$. However, it is straightforward to obtain that for ground state with zero-temperature and zero-potential limit, the near horizon geometry is $AdS_2\times \mathbf{R}^2$. The conformal symmetry in the boundary theory is still a challenge. However, it has been addressed in \cite{Blake:2013bqa} that massive gravity may be treated as a low-energy bulk description of some system with momentum dissipation, and the effort of deriving holographic massive gravity from general relativity
in AdS has been made in \cite{Kiritsis:2006hy,Aharony:2006hz,Apolo:2012gg}. }
\begin{eqnarray}
ds^2&=&-f(r)dt^2+\frac{dr^2}{f(r)}+r^2(dx^2+dy^2), \label{3} \\
f_{\mu\nu}&=&\mathrm{diag}(0,0,c_0^2h_{ij}), \label{4}
\end{eqnarray}
with $\mathcal{U}_1=2c_0/r$, $\mathcal{U}_2=2c_0^2/r^2$, and $\mathcal{U}_3=\mathcal{U}_4=0$
and
\begin{eqnarray}
 f(r)&=&\frac{r^2}{l^2}-\frac{m}{r}+\frac{q^2}{r^{2}}+\frac{c_0c_1m_g^2}{2}r+c_0^2c_2m_g^2\nonumber\\
 &=&\frac{8\pi P}{3}r^2-\frac{m}{r}+\frac{q^2}{r^{2}}+\frac{c_0c_1m_g^2}{2}r+c_0^2c_2m_g^2,\label{6}
\end{eqnarray}
where in the second line, we have used the definition of pressure \eqref{eq-Pl} and set the volume of two dimensional space to be $1$. The extended thermodynamics of massive gravity has been studied in \cite{Xu:2015rfa,Zou:2017juz}.  The integral constant $m$ and $q$ are connected with the mass and charge of the black hole as $M=m/8\pi$ and $Q=q/8\pi$, respectively. Then the requirement of $f(r_h)=0$ gives us
\begin{eqnarray}
M&=&\frac{c_0 c_1 m_g^2 r_h^2}{16 \pi }+\frac{c_0^2 c_2 m_g^2 r_h}{8 \pi }+\frac{P r_h^3}{3}+\frac{q^2}{8 \pi  r_h},\label{7}
\end{eqnarray}
which is the mass of black hole connected with the energy density of thermodynamical system as shown in \cite{Blake:2013bqa}.
The Hawking temperature $T$, the entropy $S$, the thermodynamic volume $V$, and the electric potential $\Phi$ were derived as
\begin{eqnarray}
T&=&\frac{1}{4\pi}f'(r_h)=\frac{c_0^2 c_2 m_g^2}{4 \pi  r_h}+\frac{c_0 c_1 m_g^2}{4 \pi }+2 P r_h-\frac{q^2}{4 \pi  r_h^3}, \label{8}\\
S&=&\int_0^{r_h}\frac{1}{T}\left(\frac{\partial M}{\partial
r}\right)_{Q,P,c_i}dr=\frac{r_h^2}{4},  \label{9}\\
V&=&\left(\frac{\partial M}{\partial
P}\right)_{S,Q,c_i}=\frac{r_h^{3}}{3},  \label{10}\\
\Phi&=&\left(\frac{\partial M}{\partial
Q}\right)_{S,P,c_i}=\frac{16\pi}{r_h}Q=\frac{2q}{r_h}.  \label{11}
\end{eqnarray}
We note that the formulas of the entropy and the thermodynamical volume are the same as those \eqref{eq-S} and \eqref{eq-V} in Einstein-Maxwell-scalar theory we showed in previous sections. By treating the coupling constants $c_i$ as thermodynamical variables, it is straightforward to obtain the generalized first law of the black hole in the extended phase space\cite{Xu:2015rfa}\footnote{We note that the study of the normal thermodynamics in massive gravity  via Hamiltonian method can be found in \cite{Cai:2014znn}.}
\begin{eqnarray}\label{firstlaw}
\mathrm{d}H&=&TdS+Vd P+\Phi dQ+\frac{c_0m_g^2r_h^2}{16\pi}dc_1+\frac{c_0^2m_g^2r_h}{8\pi}dc_2.\label{12}
\end{eqnarray}
Then, we look at the scaling (i.e. the dimensions) of the various thermodynamical quantities, see for instance \cite{Kastor:2009wy}. Here we have
$M,Q\propto [Length]^1,S\propto [Length]^2,P\propto [Length]^{-2},c_1\propto [Length]^{-1}$ and $c_2 \propto [Length]^0$, therefore the modified Smarr relation of the black hole is
\begin{eqnarray}
1\cdot M=H&=&2\cdot TS-2\cdot \left(\frac{\partial M}{\partial
P}\right) P+1\cdot \left(\frac{\partial M}{\partial
Q}\right) Q+(-1)\cdot \left(\frac{\partial M}{\partial
c_1}\right)c_1+0\cdot \left(\frac{\partial M}{\partial
c_2}\right)c_2
\nonumber\\
&=&2TS-2VP+\Phi Q-\frac{c_0c_1m_g^2}{16\pi}r_h^2, \label{13}
\end{eqnarray}
where the term with $c_2$ does not contribute because the dimension is zero.

\subsection{Joule-Thomson expansion}
We continue to study the Joule-Thomson expansion in massive gravity.  Using \eqref{eq-mu}, we get the Joule-Thomson expansion coefficient in massive gravity
\begin{eqnarray}
\mu_{JT}=-\frac{2 r_h^3 \left(8 \pi  P r_h^2-12 \pi  T r_h-c_0^2c_2m_g^2\right)+6 q^2 r_h}{3 \left(r_h^2 \left(8 \pi  P r_h^2+c_0c_1m_g^2 r_h-c_0^2c_2m_g^2\right)-q^2\right)}.
\end{eqnarray}
Using $\mu_{JT}=0$ we find the inversion temperature
\begin{eqnarray}\label{eq-Ti3}
T_i=\frac{8 \pi  P_i r_h^4-c_0^2c_2m_g^2 r_h^2+3 q^2}{12 \pi  r_h^3}.
\end{eqnarray}
Making use of the Hawking temperature, we obtain that for the inversion points, the horizon should satisfy the equation
\begin{eqnarray}\label{eq-equTP}
8 \pi  P_i r_h^4+\frac{3}{2} c_0c_1m_g^2 r_h^3+2 c_0^2c_2m_g^2 r_h^2-3 q^2=0.
\end{eqnarray}
We can solve $r_h$ from the above equation and then substitute it into \eqref{eq-Ti3} to get the $T_i$ as a function of $P_i$.  In what follows, we shall fix $c_0=1$ without loss of generality and mainly study the effects of $c_1$ and $c_2$.
\begin{figure}[h]
\center{
\includegraphics[scale=1]{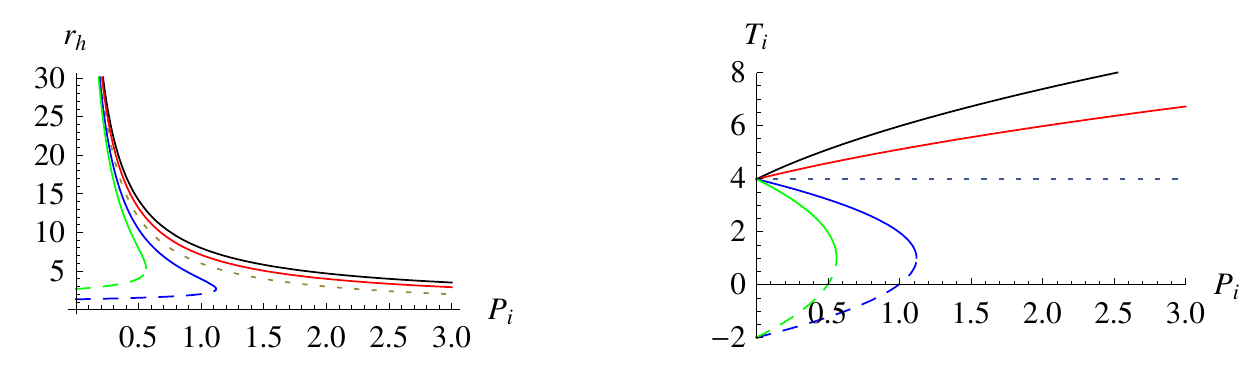}
\includegraphics[scale=1]{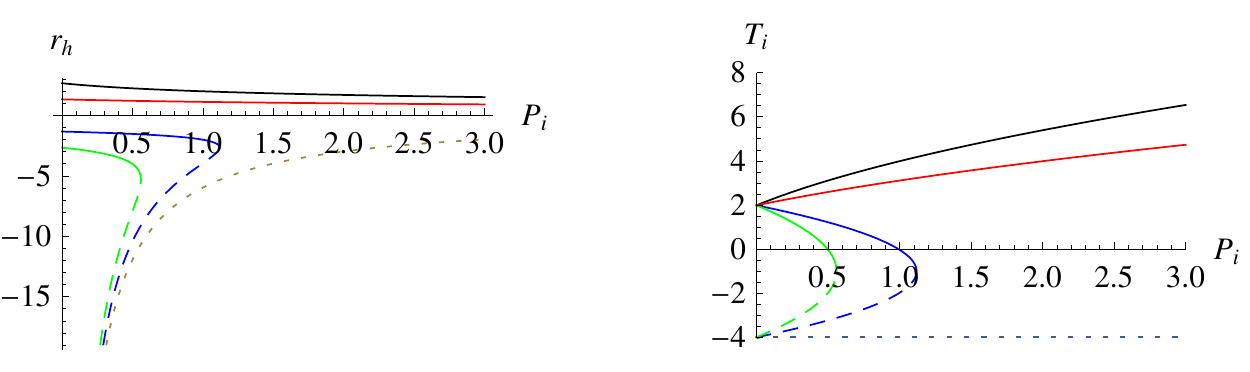}
\caption{\label{fig-Pi-Ti-q0-c2}Left: Positive horizon $r_h$ for the inversion curve. Right: Inversion curves $T_i-P_i$ for heating and cooling processes for different values of the couplings. The upper plots are for $c_1=-1$ while the bottom plots are for $c_1=1$. Colors denote different $c_2$ with  $c_2=2$ (green), $c_2=1$ (blue), $c_2=0$ (dashed), $c_2=-1$  (red) and $c_2=-2$ (black).}}
\end{figure}
\begin{figure}[h]
\center{
\includegraphics[scale=1]{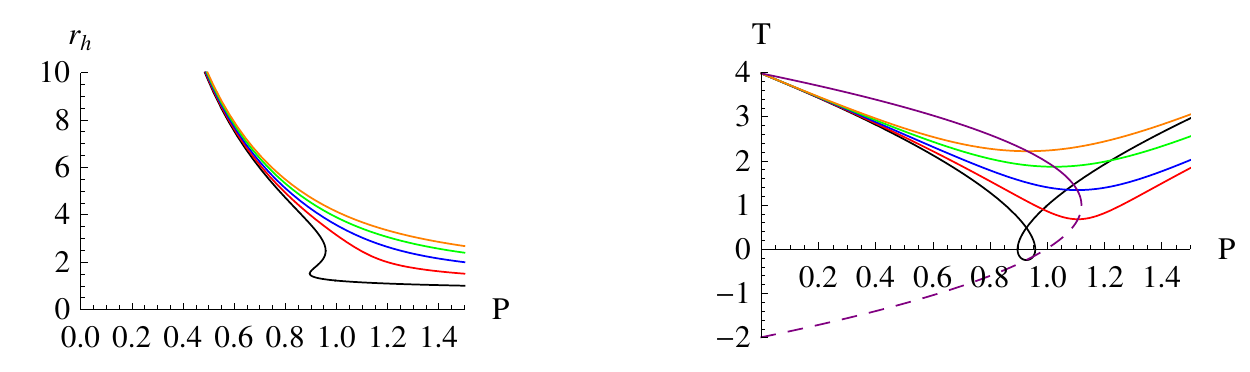}
\caption{\label{fig-P-T-q0-c2a}Left: The positive $r_h$ for the Isenthalpic curves for $c_1=-1$ and $c_2=1$. Right: Isenthalpic curves for heating and cooling processes for different values of the couplings.}}
\end{figure}

We firstly consider the case with $q=0$. Consequently, beside $r_h=0$, there are two more solutions of \eqref{eq-equTP} which are
\begin{equation}
r_h=\frac{\pm\sqrt{9 c_1^2 m_g^4-256 \pi  c_2 m_g^2 P_i}-3 c_1 m_g^2}{32 \pi  P_i}.
\end{equation}
The inversion curves and the related horizon are shown in figure \ref{fig-Pi-Ti-q0-c2}. In the upper plots, we set $c_1=-1$ to draw the positive $r_h$ (left) satisfying $\mu_{JT}=0$ and the related inversion curve (right). We observe two branches in the inversion curve for $c_2=2$ and $c_2=1$, and the related horizons are both positive. Furthermore, we study the isenthalpic curves  for $c_1=-1$ and $c_2=1$ in the right plot of figure \ref{fig-P-T-q0-c2a} where the purple line is the inversion curve\footnote{Results for $c_1=-1$ and $c_2=2$ are similar.}. From the orange line to the black one, the mass of the black hole are $M=6,5,4,3.2$ and $2.5$ while the left plot is for the related  horizon which are all positive.
In the bottom plots, we draw the inversion curve for $c_1=1$ at the right side and at the left side we show the corresponding horizon. We see that for $c_2=2$ and $1$, even though there are two branches for the inversion curves, the horizons are all negative. So these cases are not physical and only the branches with $c_2=-2$ and $c_1=-1$ are physically significant.

These observations are novel in the following aspects. Comparing with previous works \cite{Okcu:2016tgt,Ghaffarnejad:2018exz,Mo:2018qkt,Lan:2018nnp,Okcu:2017qgo,Mo:2018rgq,Kuang:2018goo}, where only one branch was obtained, we first obtain two branches for the inversion curve, which is similar to the Van der Waals fluids case. However, the minimal inversion temperature is negative unlike the case of Van der Waals fluids. On the other hand, for the isenthalpic curves in figure \ref{fig-P-T-q0-c2a}, the points $\mu_{JT}=(\partial T/\partial P)_M=0$ all fall in the inversion curve. However, we see here that $\mu_{JT}=0$ denotes the minimum but not maximal value which is a different behavior respect to the van der Waalss case and the other black hole cases that have been analyzed. This means that in the left side of the inversion curve, the isenthalpic process is a cooling process because of $\mu_{JT}<0$ while it is a warming process with $\mu_{JT}>0$ in the right side.
\begin{figure}[h]
\center{
\includegraphics[scale=0.7]{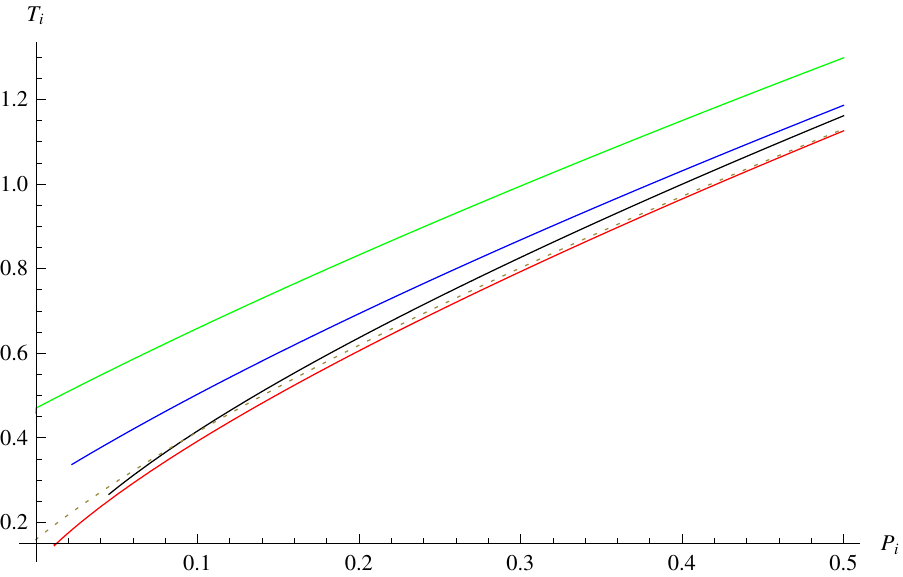}
\caption{\label{fig-Pi-Ti-q5-c2}Inversion curves $P_i-T_i$ for heating and cooling processes with $c_1=1$. From bottom to up, $c_2$ are  $(-4,-2,0,2,4)$.}}
\end{figure}
\begin{figure}[h]
\center{
\includegraphics[scale=1.2]{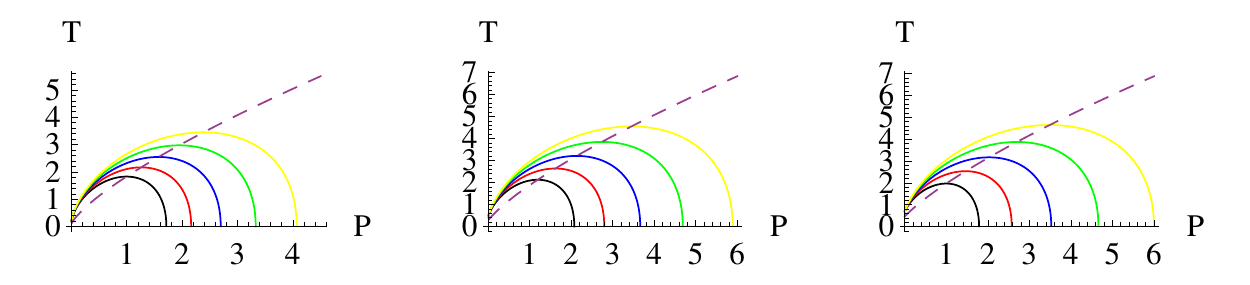}
\caption{\label{fig-P-T-q0-c2b} Isenthalpic curves for heating and cooling processes with $c_2=-2,2,4$.}}
\end{figure}

We then consider the case with $q=5$. In this case, there are four solutions of  equation \eqref{eq-equTP} for $r_h$ , which we do not show due to esthetics purposes. We find with samples of $c_1$ and $c_2$, that only one solution is real and positive, therefore getting one branch for the inversion curve. We show the inversion curves with $c_1=1$ for different $c_2$ in figure \ref{fig-Pi-Ti-q5-c2}, which is similar to the ones previously obtained in the literature with only one branch. Then the isenthalpic curves and the related inversion curves for choices of $c_2$ are shown in figure \ref{fig-P-T-q0-c2b}. Similarly, in each plot, the isenthalpic process in the left side of the inversion curve denotes warming process while those in the right side are for cooling process. Similar properties can be obtained for $c_1=-1$.

\section{Final comments}


By considering the cosmological constant as a thermodynamical quantity we have analyzed the Joule-Thomson expansion, this means, the expansion of gas from a higher pressure section to a lower one by maintaining the enthalpy of the process constant, this in the context of AdS planar black hole that exhibit momentum dissipation.
Real materials relax momentum, behavior that from the point of view of gravitational dual theories, can be introduced by several methods that break translational invariance in the field theory side. Two methods were investigated, first when linear scalar fields, massless scalar fields that depend linearly on the horizon coordinates, are introduced, and secondly the case in which the Einstein-Hilbert action is supplemented with massive potentials that renders gravity massive and that break the bulk diffeomorphism invariance of the theory. By studying the Joule-Thomson coefficient, $\mu_{JT}$, which determines the transition from warming/cooling phases, we have computed the inversion curves in the $T_i-P_i$ plane as well as the corresponding isenthalpic curves.

We have observed that for the case of linear scalars, when they possess standard kinetic term, the inversion curve possesses only one branch, similar to what was obtained in \cite{Okcu:2016tgt,Ghaffarnejad:2018exz,Mo:2018qkt,Lan:2018nnp,Okcu:2017qgo,Mo:2018rgq,Kuang:2018goo}, behavior that differs from the case of van der Waals fluids. The net effect of the momentum relaxation mechanism, which is controlled by our coupling $\beta$, is that the inversion curve is enlarged for higher values of $\beta$. This means that the temperature for which the heating/cooling transition takes place is greater when increasing $\beta$. In fact, the momentum relaxation parameter behave as an electric charge, not only enhancing the inversion curve, but also supporting the Joule-Thomson expansion in the absence of electric charge.

Next, we have modified the kinetic term for our scalars by including a nonlinear kinetic term of the type $(\partial_\mu\psi\partial^\mu\psi)^k$, contribution controlled by the parameter $\gamma$. We observe a similar behavior than in the previous case, this means, that considering greater values of $\gamma$ we obtain enlarged inversion curves. Nevertheless we observe (for the $k=2$ case) that allowing $\gamma$ to be negative, this means, by considering possible phantom contributions the inversion curve presents two branches. However, with the same parameters, the mass of black hole is negative which implies that in this case, the Joule-Thomson expansion breaks down for negative $\gamma$ because negative $\gamma$ involves instability.

For the case of the massive gravity theory we have found an interesting new behavior of the process, mostly related with the form of the isenthalpic curves. In the uncharged case, we observe that for some values of the relevant parameters we obtained two branches, but the related  horizons and the temperature are all negative which are not physically significant. Nevertheless, when constructing the isenthlapic curves we observe that the inversion point represent a minimum of the isenthalpic curve instead of a maximum as it has been typically found \cite{Okcu:2016tgt,Ghaffarnejad:2018exz,Mo:2018qkt,Lan:2018nnp,Okcu:2017qgo,Mo:2018rgq,Kuang:2018goo}. This implies that in the left side of the inversion curve, the isenthalpic process is a cooling process while it is a warming process in the right side.
So far black holes where found to always cool when passing the inversion curve, nevertheless these solutions of massive gravity are able to heat when crossing it. The situation is restored to typical behaviors when including electric charge. We note that in the present work, we mainly focus on  the possible
features of the JT expansion. It would be very interesting to scan the values of parameters in accompany with the PV criticality \cite{Xu:2015rfa,Zou:2017juz} and further
study the thermodynamical phenomenon in massive gravity. We will leave it for future work.

\section*{Acknowledgements}
 We appreciate the anonymous referee whose advice has improved the quality of this
manuscript significantly. A. C.'s work is supported by Fondo Nacional de Desarrollo Cient\'ifico y Tecnol\'ogico Grant No. 11170274 and Proyecto Interno Ucen I+D-2016, CIP2016. X.M.Kuang is supported by the Natural Science Foundation of China under Grant No.11705161 and Natural Science Foundation of Jiangsu Province under Grant No.BK20170481.

\end{document}